\input{aipcheck}

\documentclass[
    ,final            
  ]
  {aipproc}

\layoutstyle{6x9}
\usepackage{graphicx}
\usepackage{epstopdf}
\newcommand{\as}{\alpha_s}
\newcommand{\un}{\underline}

\begin{document}

\title{Development of Chaos \\in the Color Glass Condensate}


\author{Kirill Tuchin}{address={
 Nuclear Theory Group,\\ Physics Department, \\ Brookhaven National Laboratory,\\
Upton, NY 11973-5000, USA}
}

\begin{abstract}
 Noting that the number of gluons in the hadron wave function is discrete, and their 
formation in the chain of small $x$ evolution occurs over discrete rapidity intervals of $\Delta y \simeq 1/\as$, we formulate the discrete version of the Balitsky--Kovchegov evolution equation and show that its solution behaves chaotically in the phenomenologically interesting kinematic region. 
 
 \end{abstract}

\maketitle


\section{Evolution as a discrete process}

The color field of an ultra--relativistic hadron is a quasi-classical non-Abelian Weizsacker-Williams field \cite{MV,Kov}. It emerges when the occupation number of the bremsstrahlung gluons emitted at a given impact parameter exceeds unity and eventually saturates at $\sim 1/\as$.  It has been argued that in a big nucleus, such that $\as A^{1/3}\gg1$,  and not very high energies the mean-field treatment is a reasonable approximation to the evolution equations.

As a result of broken scale symmetry of QCD there exist the dimensional scale $\Lambda$ which is the infrared cutoff on the gluon's momenta. An introduction of an infrared cutoff $\Lambda$ on the momentum of the emitted gluons amounts to imposing the boundary condition. This is equivalent to the quantization 
of the gluon modes in a box of size $L \sim \Lambda^{-1}$, in which case the spectrum of the emitted gluons and their number become discrete.   The formation of a gluon occurs over a rapidity interval of $\Delta y \simeq 1/\as$. Therefore, the evolution in rapidity can be considered as a discrete quantum process, where each subsequent step occurs when $\Delta y\, \as \simeq 1$ \cite{chaos}.

Assuming that $\Delta y\,\as$ is a certain number for all steps in evolution process neglects a stochastic nature of quantum evolution.  Full treatment of the discrete BK equation requires taking these effects into account. However, unfortunately BK equation is known to resist all attempts of analytical solution, and our hope at present is to develop a meaningful approximation. Thus, we suggest an approximation in which the gluons are emitted over a fixed ``time" defined by $\bar\alpha_s\Delta y =C$ with $C=1$.
To justify this assumption, let us note that BFKL takes into account only fast gluons, i.e.\  those with $C\sim 1$. It is beyond the leading logarithmic (LL) approximation to take into account slow gluons. Moreover, it is known that an account of NLL corrections effectively leads to imposing a rapidity veto \cite{veto} on the emission of gluons with close rapidities, which restricts production of gluons with small $C$ (this is due to an effective repulsion between the emitted gluons induced at the NLL level). Therefore $C$ is bounded from below by a number close to one. On the other hand the probability that no gluon is emitted when $C$ becomes larger than one is very small if we choose the high density initial condition, such as the one given by the McLerran-Venugopalan model\cite{MV}.  Therefore, $C$ takes random values around 1, but the effective dispersion can be expected quite small.  

\section{Discrete evolution equation}

\begin{figure}\label{fig1}
  \includegraphics[height=5cm]{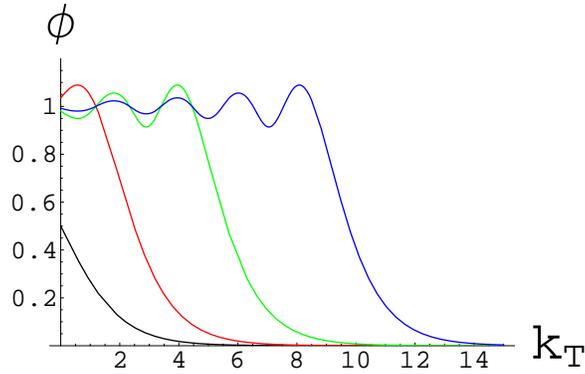}
  \caption{Discrete BK equation at $\omega=2.8$. Different lines from left to right correspond to $n=1,4,7,10$ (black, red, green, blue).}
\end{figure}

\begin{figure}\label{fig2}
  \includegraphics[height=5cm]{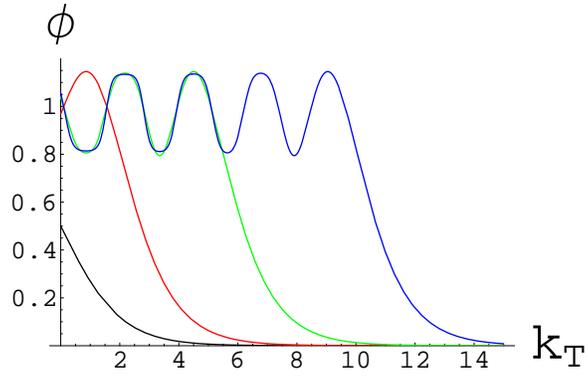}
  \caption{Discrete BK equation at $\omega=3.1$. Different lines from left to right correspond to $n=1,4,7,10$ (black, red, green, blue).}
\end{figure}

\begin{figure}\label{fig3}
  \includegraphics[height=5cm]{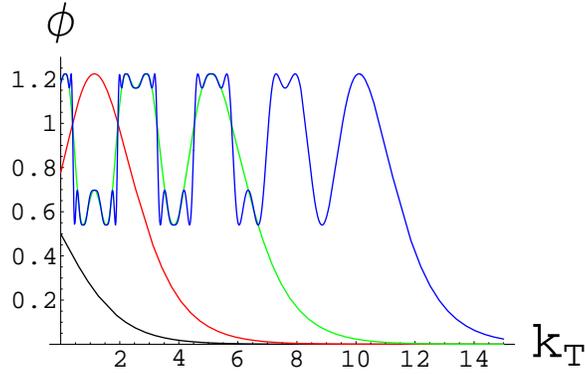}
  \caption{Discrete BK equation at $\omega=3.495$. Different lines from left to right correspond to $n=1,4,7,10$ (black, red, green, blue).}
\end{figure}

Equation which describes the gluon evolution in the high parton density regime of QCD is the Balitsky-Kovchegov equation \cite{Balitsky:1995ub,Kovchegov:1999yj}.  This equation is formulated for the forward scattering amplitude of a color dipole of transverse momentum $\un k$ and rapidity $y$ to scatter of a big nucleus at impact parameter $\un b$.  Assumption that the dispersion of the dipole transverse momenta with $y$ is a slow process one can develop  a diffusion approximation to the BK equation.  This approximation is meaningful in the saturation region. The discrete version of BK equation takes the form
(see also \cite{Bialas:2005md})
\begin{equation}\label{desK}
\tilde N_{n+1}(\un k,y)\,=\,(1\,+\,\chi(\gamma_0))\,\tilde N_n(\un k,y)\,- \,\tilde N_n^2(\un k,y)\,,
\end{equation}
where we introduced the discrete variable $n$ to enumerate emitted gluons and $\chi(\gamma_0)=4\ln2\equiv \omega-1$. It is convenient to re-scale the scattering amplitude $\phi_n= \chi(\gamma_0)\tilde N_n$ so that the corresponding continuous amplitude is normalized to $\tilde N_n(\un k,y)\le 1$:
\begin{equation}\label{phi}
\phi_{n+1}\,=\,\omega\,\phi_n\,-\,(\omega\,-\,1)\phi_n^2\,.
\end{equation}
 Numerical solution to the discrete BK equation is shown in Figures \ref{fig1}--\ref{fig4}.

Let us now consider how does the evolution proceeds for various values of $\omega$.
In Fig.~\ref{fig1} the case of $1<\omega<3$ is shown. In that case, for any $k$ 
there is one stable fixed point at $\phi_n=1$ and one unstable fixed point at $\phi_n=0$.
The fixed points are determined from the condition $\phi_{n+1}=\phi_n$. 

In Fig.~\ref{fig2} we consider the case $3<\omega<3.442\ldots$.  The point $\phi_n=1$ ceases to be unstable. Instead two new stable points appear. These can be determined from the  condition $\phi_{n+2}=\phi_n$. 
The described multiplication of stable points is called in general \emph{bifurcation}. In our particular case it is referred to as the \emph{period doubling scenario}.  

When $3.442\ldots<\omega<3.56\ldots$ there are four new fixed points, see Fig.~\ref{fig3}.
It is important to emphasize that at $n\to \infty$ the value of $\phi_n$ settles to a given set of fixed points (specified by the value of $\omega$) independently of the initial condition. In other words, at very high energies the scattering amplitude in the saturation regime $k_T<Q_s(n)$ is independent of the initial condition. 

\section{Onset of chaos}

\begin{figure}\label{fig4}
  \includegraphics[height=5cm]{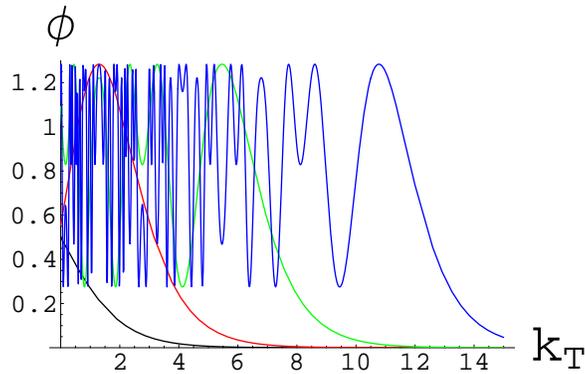}
  \caption{Discrete BK equation at $\omega=3.77$. Different lines from left to right correspond to $n=1,4,7,10$ (black, red, green, blue).}
\end{figure}

The period doubling proceeds at increasingly smaller increments of $\omega$ until the \emph{ accumulation point} $\omega_F=3.569\ldots$ known also as the \emph{Feingenbaum's number}. At this point there is no more universal limiting behavior at large $n$. Instead small change in the initial condition leads to large change in the final state. The onset of this chaotic behavior can be observed in Fig.~\ref{fig4}.  
Note now that at first, $\omega_{BFKL}=1+4\ln2=3.77>\omega_F$ and at second, $\omega_{BFKL}$ is the absolute minimum of the function $1+\chi(\gamma)$. Thus, we conclude that the evolution of the scattering amplitude at high energies in the saturation region may be chaotic. At the same time, it is evident from Figs.~\ref{fig1}--\ref{fig4} that the high $k_T$ tail of the scattering amplitude which describes the perturbative regime is not affected by the peculiar behavior of the discrete equation.   

By averaging over all events one can define the mean value of the scattering amplitude.
However, this procedure hides a lot of interesting physics. The most obvious example of this is diffraction, which measures the strength of fluctuations in the inelastic cross section. Figs.~\ref{fig1}--\ref{fig4} imply that diffraction is a significant part of the total inelastic cross section at very high energies, and is universal (independent of the properties of the target). 

The model used in this letter is admittedly oversimplified: we neglected the diffusion in transverse momentum, stochasticity of gluon emission and the dynamical fluctuations beyond the mean field approximation. Nevertheless, we hope that at least some of the features of discrete quantum evolution at small $x$ will survive a more realistic treatment. The chaotic features of small $x$ evolution 
open a new intriguing prospective on the studies of hadron and nuclear interactions at high energies.

\begin{theacknowledgments}
 This research was done in the collaboration with Dima Kharzeev. I would like to thank Yuri Kovchegov, Alex Kovner, Misha Kozlov, Genya Levin, Larry McLerran, Al Mueller and Anna Stasto for informative and helpful discussions and comments.
This research was supported by the U.S. Department of
Energy under Grant No. DE-AC02-98CH10886. BNL preprint number is BNL-NT-05/24.
\end{theacknowledgments}

\end{document}